\documentclass{moriond}

\def\be{\begin{equation}}
\def\ee{\end{equation}}
\def\bea{\begin{eqnarray}}
\def\eea{\end{eqnarray}}

\newcommand{\Photo}{\includegraphics[height=35mm]{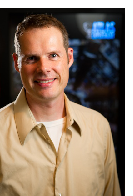}}
\newcommand{\met} {\mbox{\ensuremath{\slash\kern-.7emE_{T}}}}

\begin{document}
\vspace*{4cm}
\title{Single top quark production at the Tevatron}

\author{ R. Schwienhorst {\sl on behalf of the CDF and D0 Collaborations} }

\address{Department of Physics \& Astronomy, Michigan State University, 
567 Wilson Road\\
East Lansing, Michigan 48823, USA\\
schwier@pa.msu.edu}

\maketitle\abstracts{
The production of single-top quarks occurs via the weak interaction 
at the Fermilab Tevatron proton-antiproton collider. Single top quark events are
selected in the lepton+jets final state by CDF and D0 and in the missing transverse
energy plus jets final state by CDF.
Multivariate classifiers separate the $s$-channel and $t$-channel single-top signals
from the large backgrounds. The combination of CDF and D0 results leads to the
first observation of the $s$-channel mode of single top quark production.
The $t$-channel and single top combined cross sections have also been measured.
}

\section{Introduction}

The top quark is central to understanding physics in the Standard Model (SM) and
beyond. The study of single top quark production in particular provides unparalleled
access to the weak interaction of the top quark~\cite{Tait1:2000sh}. This paper
summarizes single-top quark measurements from the Tevatron
proton-antiproton collider at Fermilab.
The Tevatron operation ended in 2011, with CDF and D0 each collecting
10~fb$^{-1}$ of proton-antiproton data~\cite{Abazov:2005pn} at a center-of-mass
energy of 1.96~TeV. 

\begin{figure}[!h!tbp]
\includegraphics[width=0.26\textwidth]{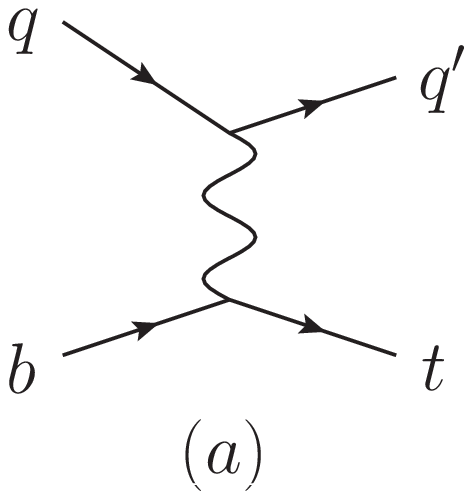}
\includegraphics[width=0.3\textwidth]{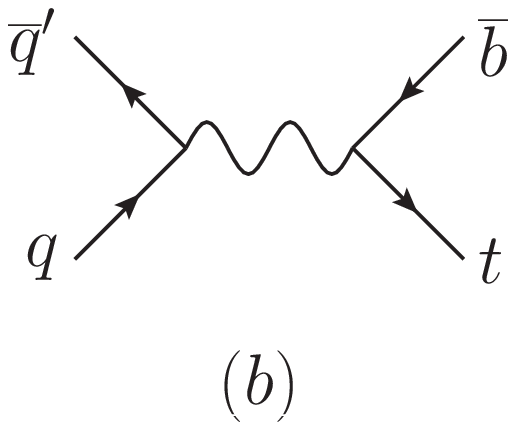}
\caption{Feynman diagrams for single top quark production in (a) the $t$-channel
and (b) the $s$-channel. 
\label{fig:feynst}}
\end{figure}

Single top quark production proceeds via the $t$-channel exchange of a $W$~boson
between a heavy quark line and a light quark line, shown in Fig.~\ref{fig:feynst}(a),
the $s$-channel production and decay of a virtual $W$~boson, shown in
Fig.~\ref{fig:feynst}(b), or the production of a top quark in association with a $W$~boson.
At the Tevatron, the $t$-channel cross section is largest, followed by the
$s$-channel, while the associated production of a top quark and a $W$~boson
is too small to be observed.

The cross section at next-to-leading order in QCD with next-to-next-to-leading-log
gluon resummation is $2.10\pm0.13$~pb for the $t$-channel~\cite{Kidonakis:2011wy}
and $1.05\pm0.06$~pb for the $s$-channel~\cite{Kidonakis:2010tc}.
These cross sections are calculated for a top quark mass
of 172.5~GeV, the same mass used in the analyses reported here, which is close to the
world average of $173.2\pm0.9$~GeV~\cite{CDF:2013jga}.

Single top quark production ($t$-channel and $s$-channel combined) was first observed
by the CDF~\cite{SGTOP-OBS-CDF} and D0~\cite{SGTOP-OBS-D0} collaborations at the
Tevatron. The $t$-channel mode was also first isolated at the Tevatron, by the 
D0 collaboration~\cite{Abazov:2011rz}. The $t$-channel mode has a large cross section
at the LHC, where it has also been observed by both ATLAS~\cite{Aad:2012ux} and
CMS~\cite{Chatrchyan:2012ep}. The $s$-channel production mode has a small cross section
at the LHC due to the quark-antiquark initial state. To date only unpublished upper
limits on $s$-channel production have been reported by the LHC collaborations.

\section{Selection and modeling}

\subsection{Event selection}\label{sec:selection}

Top quarks decay essentially always to a $W$~boson and a bottom quark. The final state is
then selected based on the decay of the $W$~boson. The lepton+jets final state has a
large signal and manageable backgrounds. The event selection requires a high-transverse
momentum isolated electron or muon, large missing transverse energy
($\met$), and two or three jets, at least one of which is required to be
identified as originating from a $b$-quark ($b$-tag). Events with misreconstructed
objects or phase space regions dominated by QCD multijet production are rejected.
About $10^5$ events are selected in data, with an expected signal contribution 
of about 400 $t$-channel events and about 250 $s$-channel events.

The CDF collaboration additionally includes the missing transverse energy ($\met$)
plus jets final state, where the decay products of the $W$~boson are not reconstructed
or include a $\tau$~lepton. This final state suffers from a large background due to
multijet production, which is reduced effectively with a neural network (NN). About
$10^6$ events are selected in data per experiment, with an expected signal contribution 
of about 250 $s$-channel events.

Events are separated into categories based on the $b$-tag information. Since $t$-channel
production results in a light quark in addition to the top quark, these events mainly
populate the single-$b$-tag category. By contrast, $s$-channel events with two 
$b$~quarks in the final state will mainly populate the two-$b$-tag category. The CDF
collaboration additionally separates events based on the $b$-tag likelihood.

\subsection{Signal and background modeling}

Simulation samples are used to model the single top signal and the dominant $W$+jets
and top pair backgrounds, as well as smaller backgrounds from $Z$+jets and diboson
production. Multijet events are modeled using data.

\subsection{Systematic uncertainties}

Sources of systematic uncertainty in the single-top measurements include the
modeling of the signal, the modeling and normalization of the backgrounds, detector
efficiencies and resolutions, jet energy scale and $b$-tag modeling. Both shape and
normalization components are included. The total uncertainty on the background is
between 15\% and 20\% depending on the analysis channel.

\section{Single top production cross-section measurement}

\subsection{CDF: combined single top}

\begin{figure}[!h!tbp]
\includegraphics[width=0.45\textwidth]{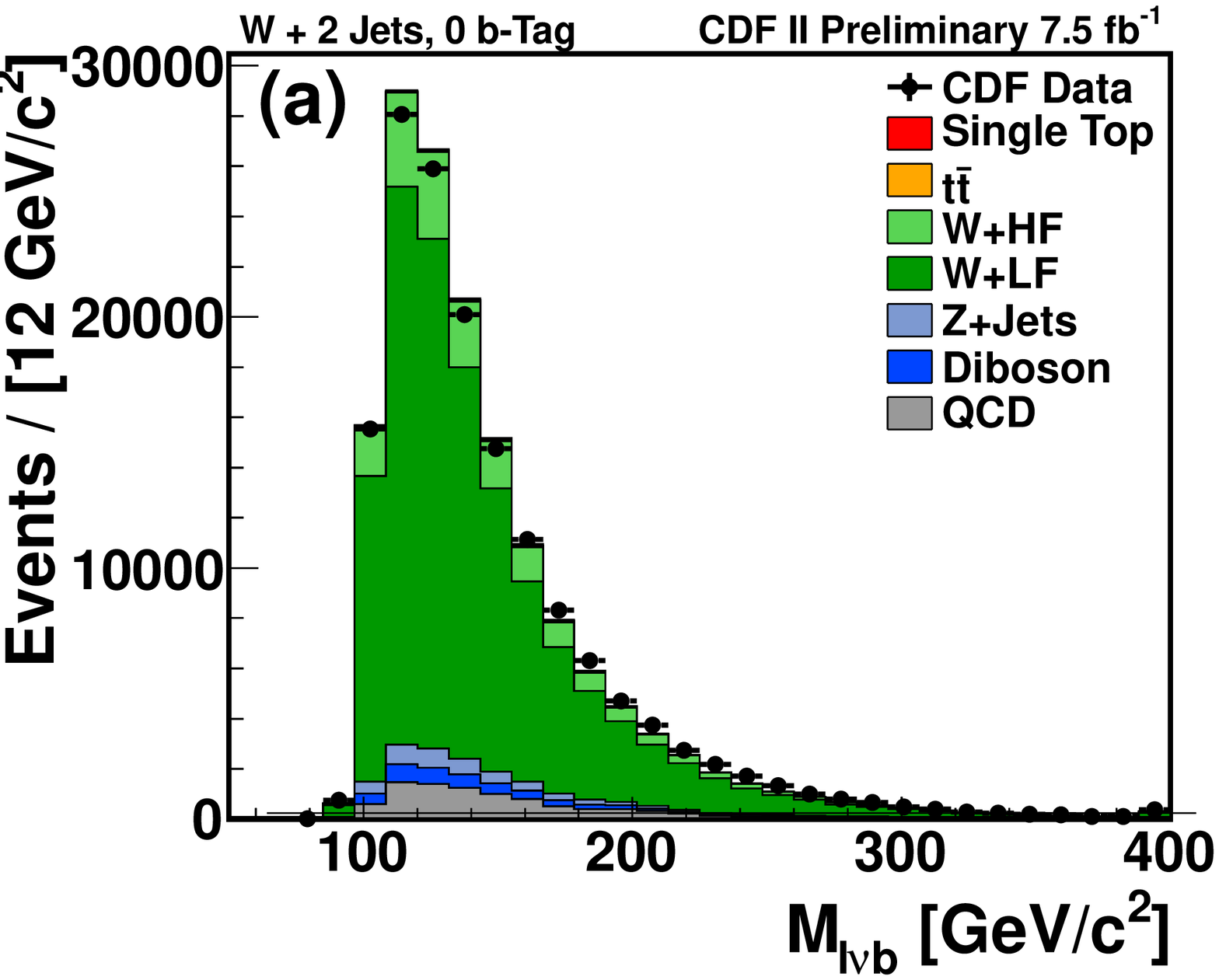}
\includegraphics[width=0.45\textwidth]{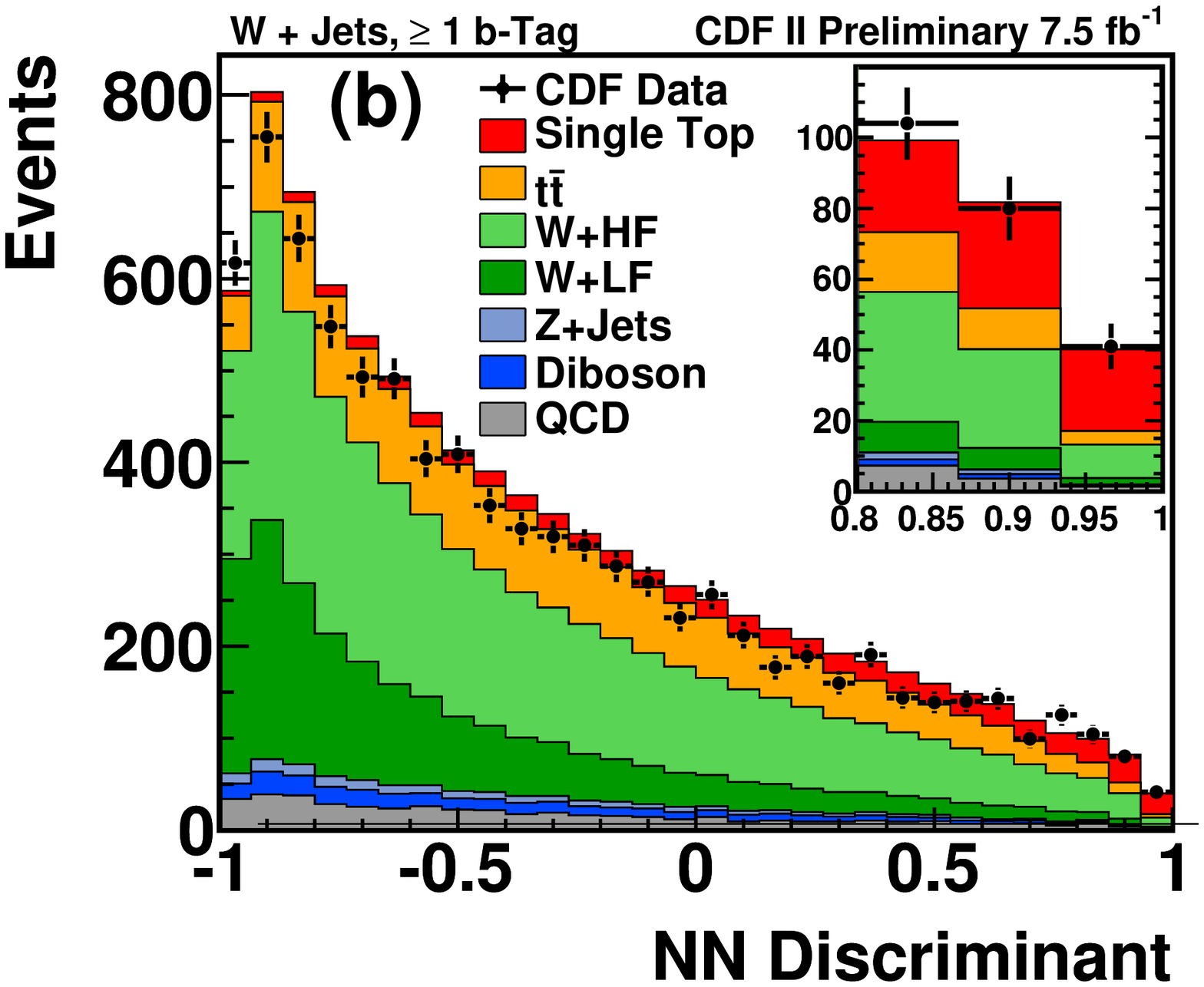}
\caption{CDF lepton+jets (a) reconstructed top quark mass for zero-$b$-tag
events and (b) NN discriminant for events with at least one $b$-tag.
\label{fig:CDFst}}
\end{figure}

The combined $t$+$s$ analysis with lepton+jets events at CDF with 7.5~fb$^{-1}$
utilizes a NN discriminant that is trained with both $t$-channel and $s$-channel
as the signal~\cite{SGTOP-CDF}. One of the main discriminating variables in the NN is
the reconstructed mass of the top quark, which is shown for a zero-tag control region
in Fig.~\ref{fig:CDFst}(a). The NN distribution in the signal region is shown in
Fig.~\ref{fig:CDFst}(b).

The single top cross section measured in lepton+jets events is 
$3.04^{ +0.57}_{ -0.53}$~pb. The cross section measured in $\met$+jets events is
$3.20^{ +1.39}_{ -1.43}$~pb~\cite{SGTOP-MJ}. The lepton+jets measurement as a function
of both $s$-channel and $t$-channel cross sections is shown in Fig.~\ref{fig:D02d}(a).

\subsection{D0: combined single top and $t$-channel}

\begin{figure}[!h!tbp]
\includegraphics[width=0.45\textwidth]{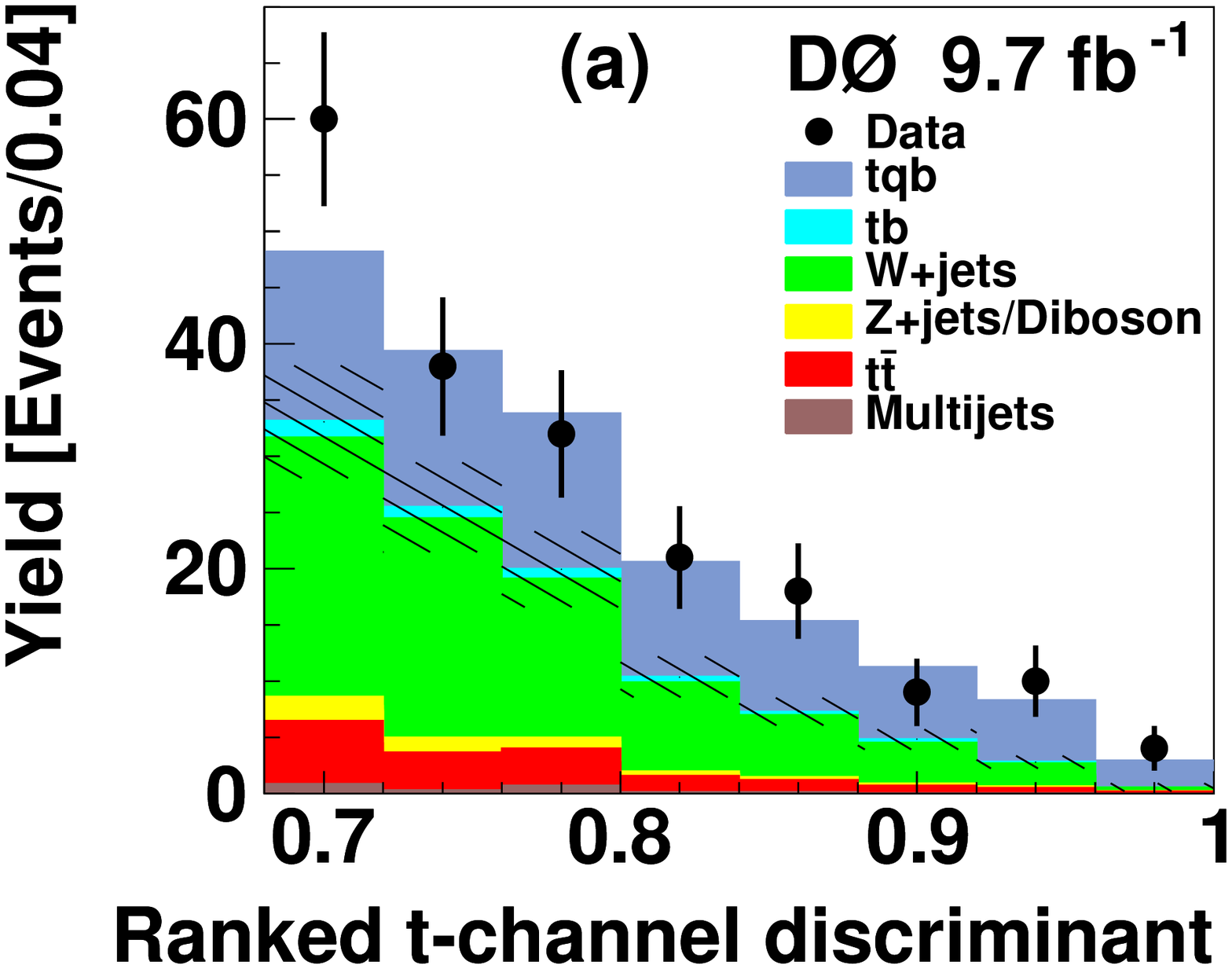}
\includegraphics[width=0.45\textwidth]{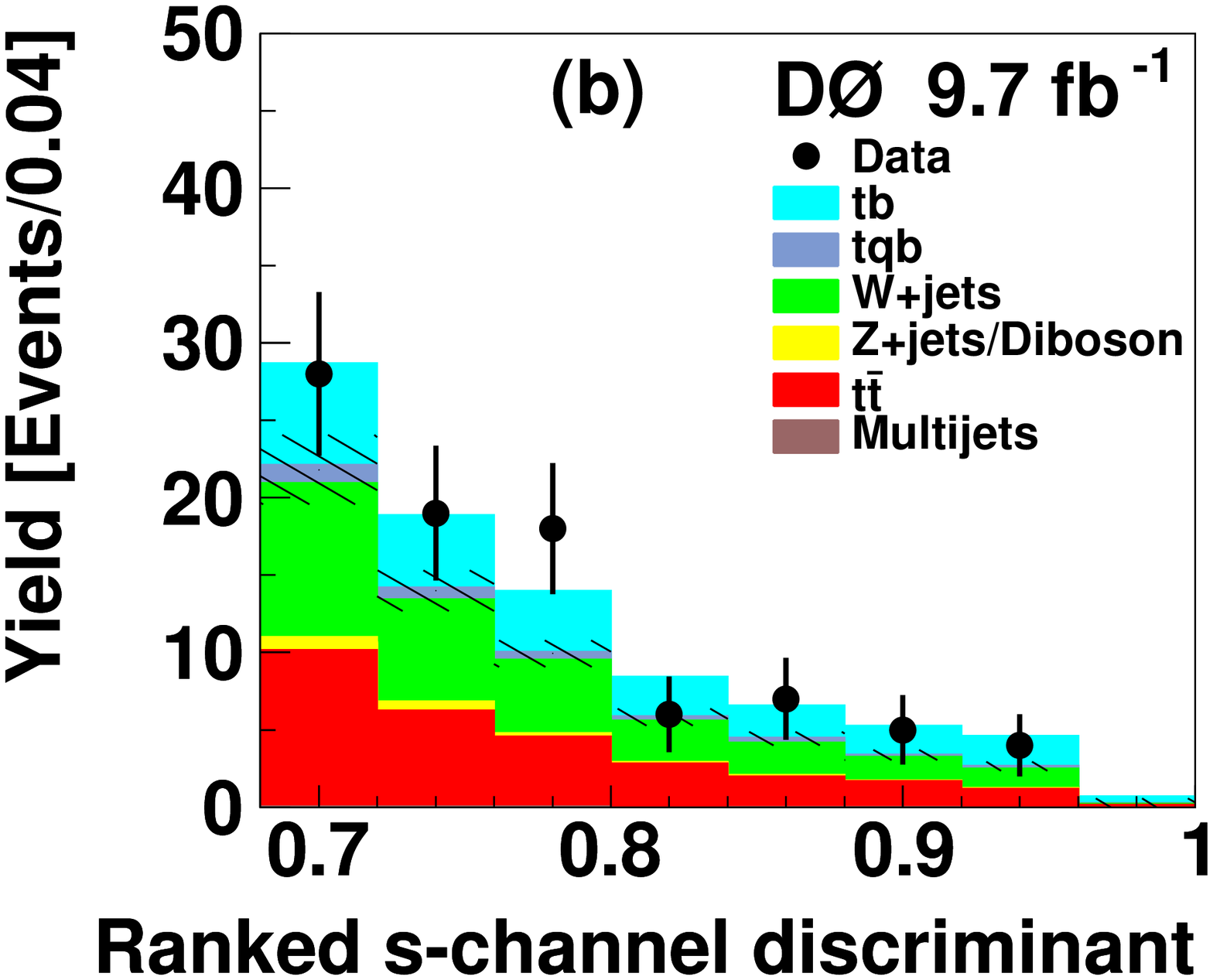}
\caption{D0 lepton+jets discriminant in the signal region (a) in the 
$t$-channel and (b) in the $s$-channel.
\label{fig:D0discr}}
\end{figure}

The D0 single top analysis forms two multivariate discriminants, one optimized for
the $t$-channel and one for the $s$-channel, shown in their respective signal regions
in Fig.~\ref{fig:D0discr}. These two discriminants are then combined into a single
discriminant that that is simultaneously
sensitive to both $t$-channel and $s$-channel production.

\begin{figure}[!h!tbp]
\includegraphics[width=0.52\textwidth]{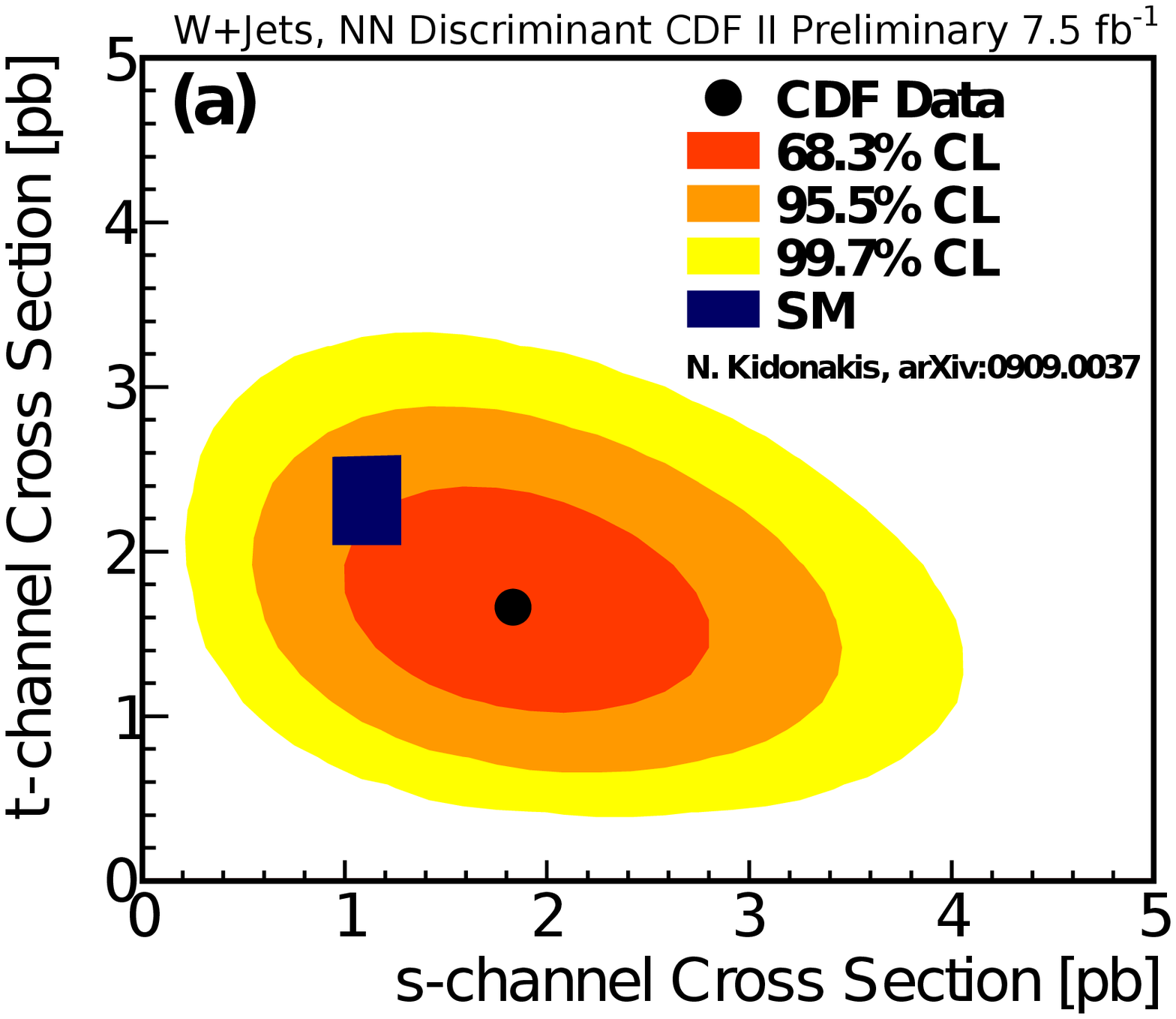}
\includegraphics[width=0.45\textwidth]{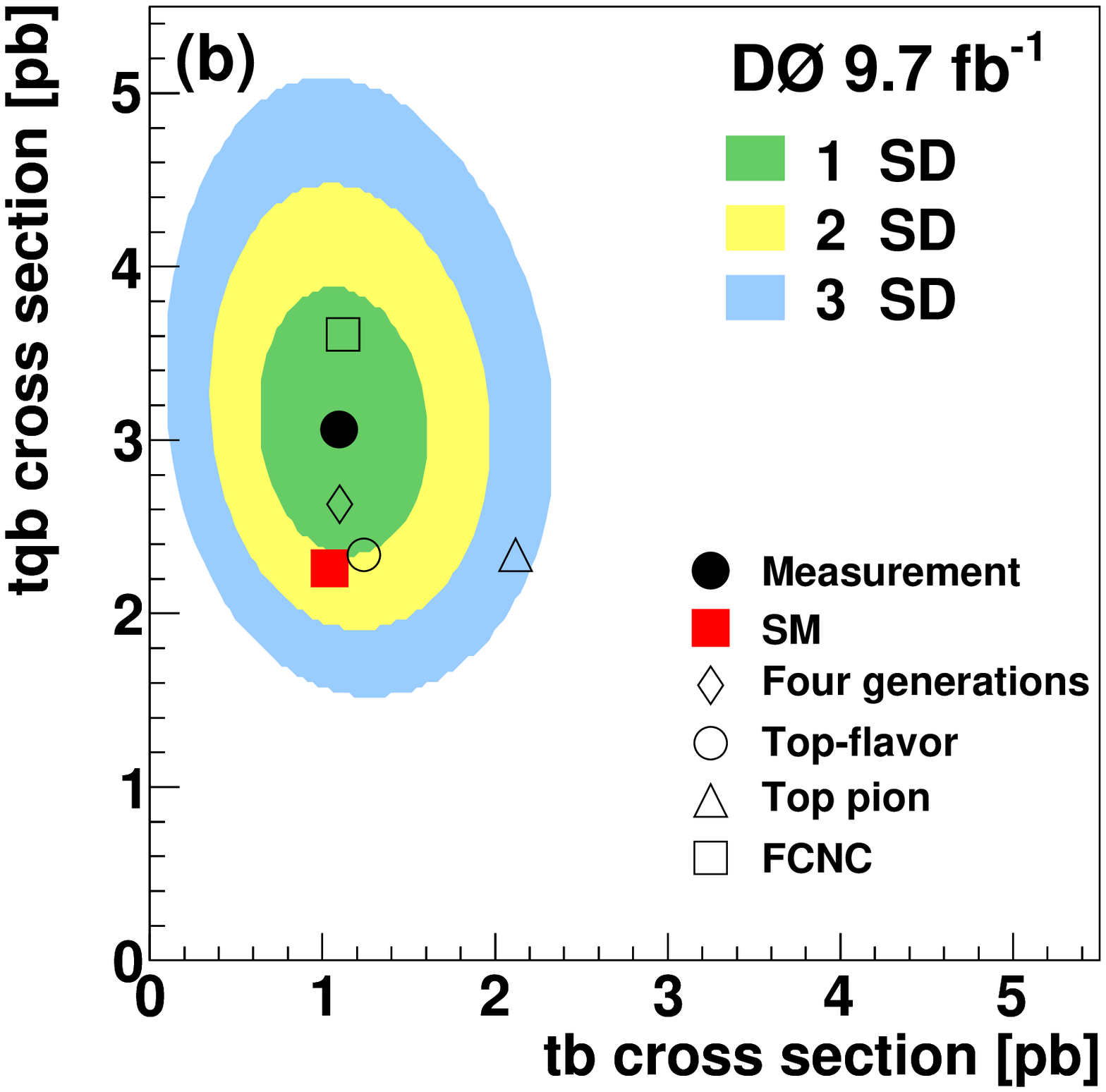}
\caption{Posterior density as a function of the $t$-channel and
$s$-channel single top cross sections for (a) the CDF lepton+jets analysis and
(b) the D0 analysis.
\label{fig:D02d}}
\end{figure}

The resulting two-dimensional posterior density distribution is shown in
Fig.~\ref{fig:D02d}(b), together with several models of new
physics~\cite{Tait1:2000sh,Alwall:2006bx,Abazov:2007ev}. The cross-section measurements
for the $t$-channel, the $s$-channel
and the combination are based on this distribution.

The $t$-channel cross section measured by D0 is $3.07^{ +0.53}_{ -0.49}$~pb
and the combined cross section is $4.11^{+0.59}_{ -0.55}$~pb.

\subsection{CKM matrix element $V_{tb}$}

Single top quark production proceeds via the $tWb$ vertex, thus the production
cross section is proportional to the square of the CKM matrix~\cite{CKM1,CKM2}
element $|V_{tb}|$. The extraction of $|V_{tb}|$ from the single top cross section
measurement does not require assumptions about the number of quark generations or
unitarity of the CKM matrix. The resulting 95\% confidence level lower limit on
$|V_{tb}|$ from the CDF lepton+jets measurement is 0.78, the lower limit from the
D0 measurement is 0.92, which is the most stringent limit from the Tevatron.

\section{$s$-channel observation}

The two-dimensional D0 posterior density distribution in Fig.~\ref{fig:D02d} is also
used to extract a cross-section measurement for the $s$-channel. The measured cross
section is $1.10^{ +0.26}_{ -0.31}$~pb, and the observed significance is 3.7 standard
deviations (SD). 

\begin{figure}[!h!tbp]
\includegraphics[width=0.68\textwidth]{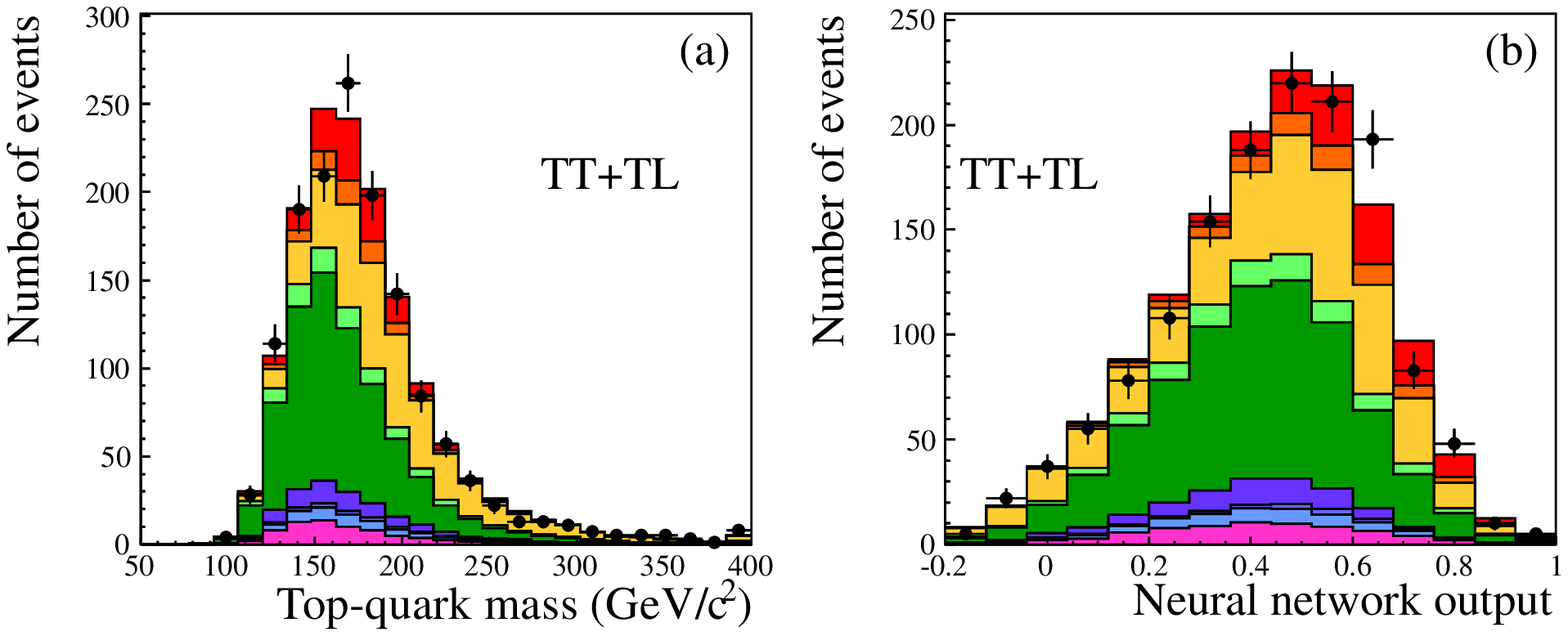}
\includegraphics[width=0.31\textwidth]{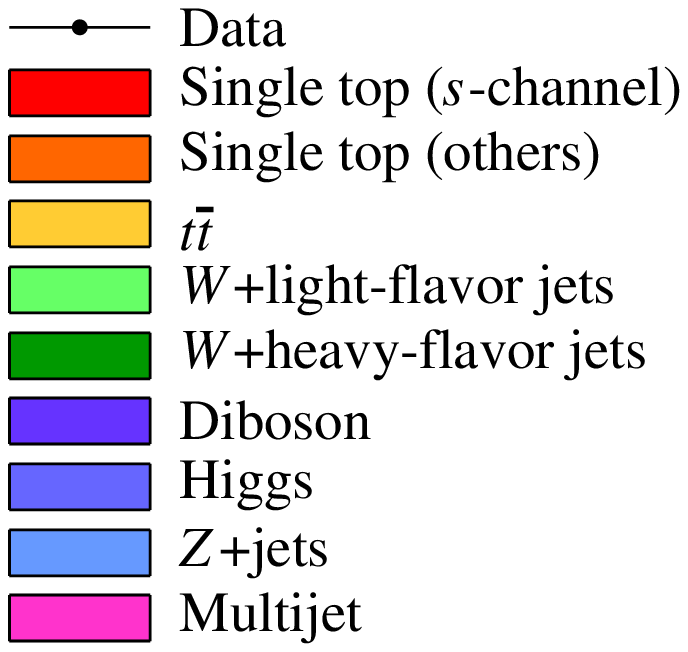}
\caption{CDF lepton+jets (a) reconstructed top quark mass and (b) $s$-channel
discriminant in events with two $b$-tagged jets.
\label{fig:CDFTT}}
\end{figure}

The CDF lepton+jets $s$-channel analysis forms a NN discriminant to separate the
$s$-channel signal from the large backgrounds in several categories of events separated
by the number loose (L) and tight (T) $b$-tagged jets~\cite{Aaltonen:2014qja}.
The discriminant distribution as well as the top quark mass are shown in
Fig.~\ref{fig:CDFTT}.

The CDF lepton+jets analysis is combined with the $\met$+jets analysis to obtain
a cross section for $s$-channel single top quark production of 
$1.36^{ +0.37}_{ -0.32}$~pb~\cite{Aaltonen:2014xta}. This corresponds to a significance
of 4.2~SD.

\begin{figure}[!h!tbp]
\includegraphics[width=0.5\textwidth]{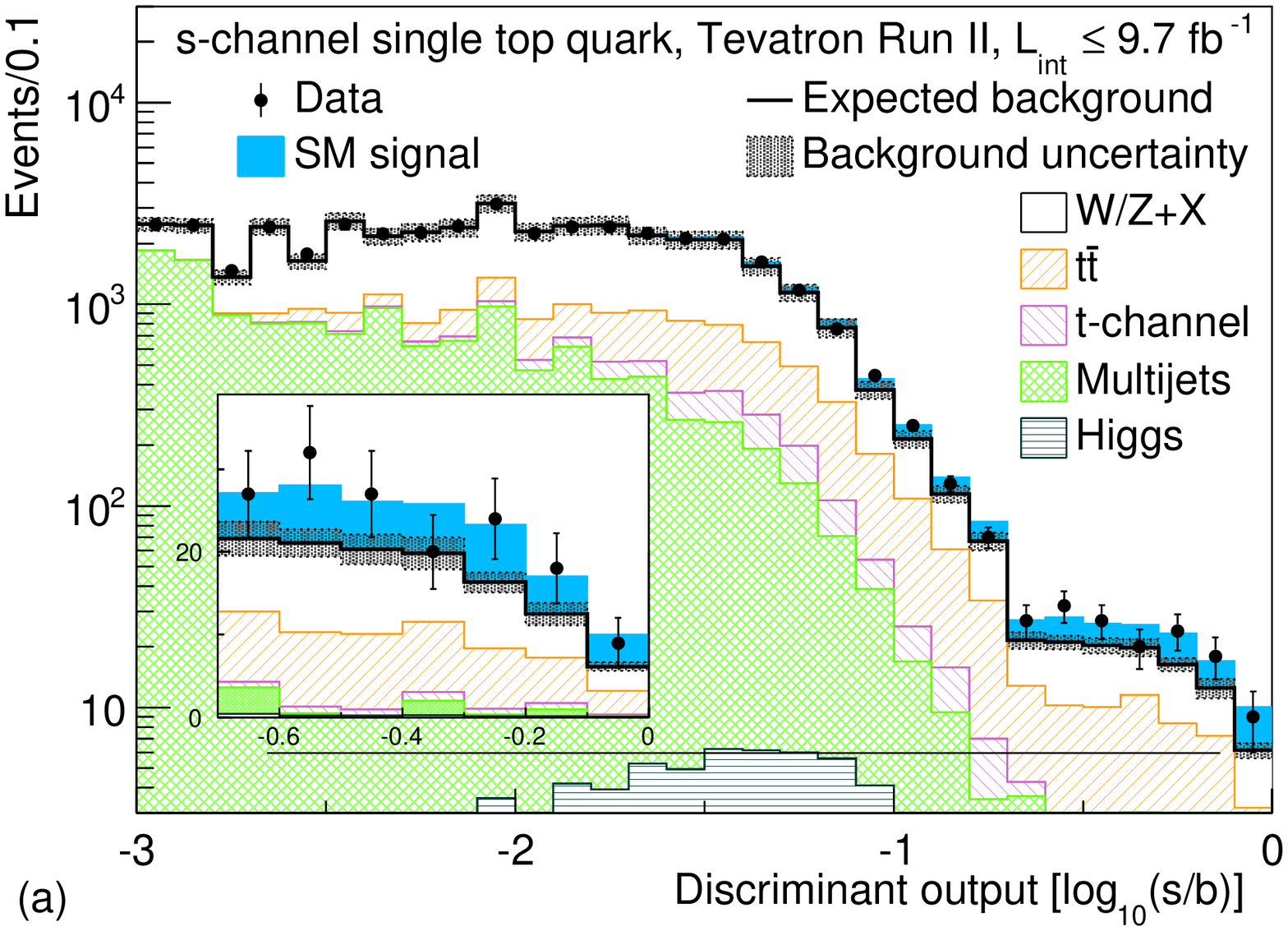}
\includegraphics[width=0.45\textwidth]{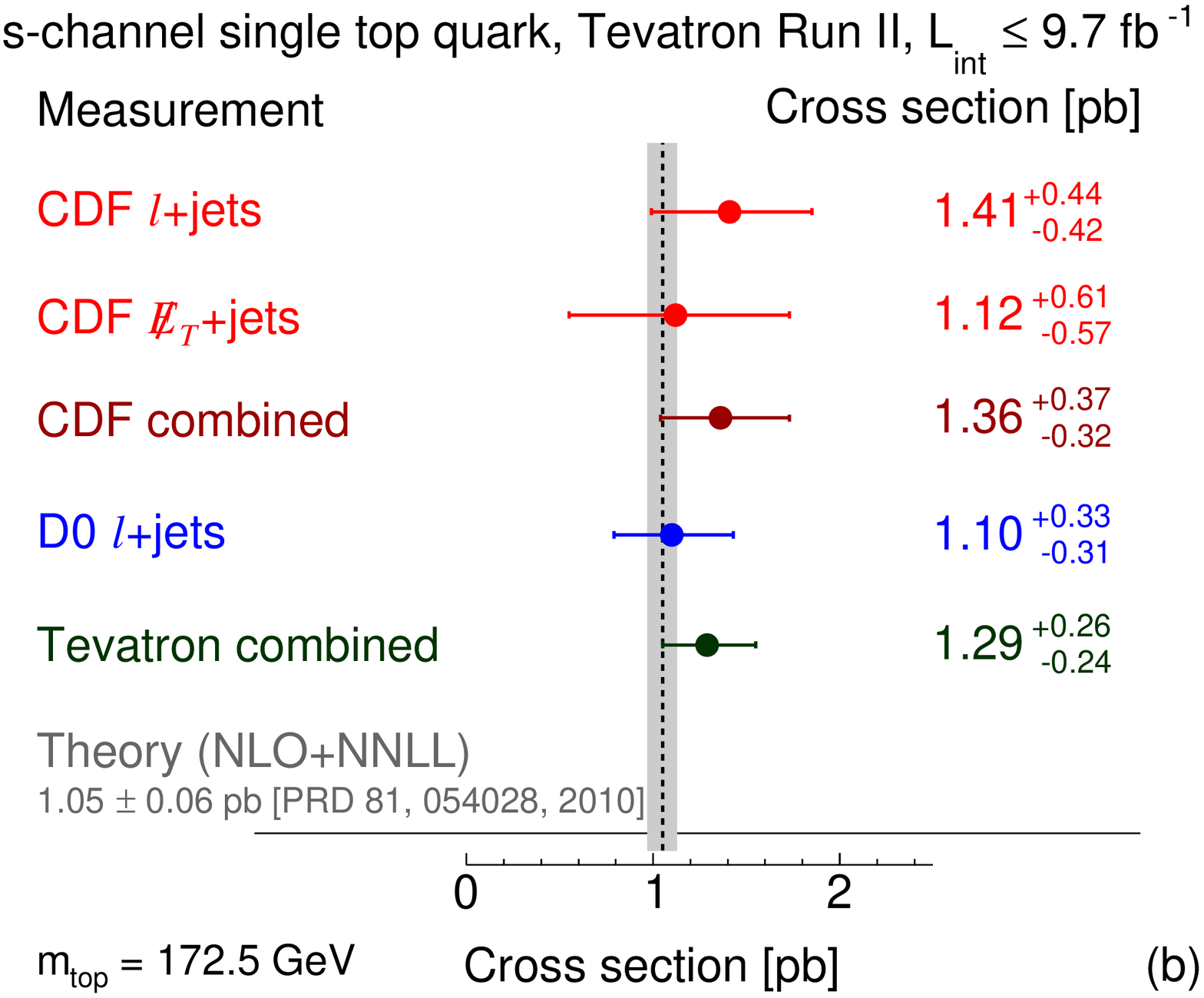}
\caption{Tevatron $s$-channel combination (a) sorted discriminant distribution and
(b) summary of results.
\label{fig:TeV}}
\end{figure}

The CDF lepton+jets and $\met$+jets discriminants are combined with the D0
discriminant in the Tevatron combination~\cite{CDF:2014uma}. The combined discriminant
is shown in Fig.~\ref{fig:TeV}(a). This combination measures a cross section for
$s$-channel production of $1.29^{ +0.26}_{ -0.24}$~pb. The measurement has a significance
of 6.3~SD, which makes this combination the first observation of single top production
in the $s$-channel, and the first observation of a process through a Tevatron
combination. A summary of all $s$-channel measurements is shown in 
Fig.~\ref{fig:TeV}(b).

\section{Conclusions}

The production of single top quarks was first observed at the Tevatron in 2009
independently by the CDF and D0 experiments. The cross section has been measured with
an uncertainty of 14\%. The two relevant production modes at the Tevatron have also
been observed separately. The $t$-channel cross section has been measured with an
uncertainty of 16\%. The $s$-channel mode
has recently been observed for the first time in a Tevatron combination of CDF and D0
with an uncertainty of 19\% and a significance of 6.3 standard deviations.

\section*{References}

\bibliography{SchwienhorstSingleTopTev}

\end{document}